# An outlook on extracellular waveforms produced by the three neuronal compartments

Jérémie Sibille[1-7*], Kai Lun Teh[5,8], Alexandra Tzilivaki[1-3], Dietmar Schmitz[1-7], Paula T. Kuokkanen[7*]

[1] Charité -Universitätsmedizin Berlin, corporate member of Freie Universität Berlin, Humboldt-Universität Berlin, and Berlin Institute of Health, Neuroscience Research Center, 10117 Berlin, Germany.
[2] German Center for Neurodegenerative Diseases (DZNE), 10117 Berlin, Germany.
[3] NeuroCure Cluster of Excellence, Charitéplatz 1, 10117 Berlin, Germany.
[4] Einstein Center for Neurosciences, Charitéplatz 1, 10117 Berlin, Germany.
[5] Bernstein Center for Computational Neuroscience, Humboldt-Universität zu Berlin, Philippstrasse. 13, 10115 Berlin, Germany.
[6] Max Delbrück Center for Molecular Medicine in the Helmholtz Association, Robert-Rössle-Straße 10, 13125 Berlin, Germany.
[7] Institute for Theoretical Biology, Humboldt-Universität zu Berlin, 10115 Berlin, Germany.
[8] Institute of Biology, Otto-von-Guericke-University Magdeburg; 39120 Magdeburg, Germany.
[*] Dual senior authorship.

Send correspondence to:
Jérémie Sibille
Neuroscience Research Center
Charité - Universitätsmedizin Berlin
Charitéplatz 1
10117 Berlin, Germany
Phone: +49 30 450 539034

Paula Kuokkanen
Institute for Theoretical Biology
Humboldt-Universität zu Berlin
Unter den Linden 6
10099 Berlin, Germany
Phone: +49 30 2093 98426

Email: jeremie.sibille@charite.de / p.kuokkanen@biologie.hu-berlin.de

# ABSTRACT

The brain is composed of billions of neurons with virtually endless morphologies and ion channel compositions, resulting in unique extracellular waveforms. Nevertheless, almost all neuronal morphologies can be reduced to a simple architecture made of three principal compartments: 1) the soma and nearby axonal hillock, 2) axonal projections ending in arbors or single synaptic contacts, and 3) dendrites. This review offers a perspective on how these three ubiquitous neuronal compartments can be identified and how they shape the extracellularly recorded waveforms, when spatial considerations are taken into account. This outlook utilizes biophysical modelling to complement existing experimental observations. Modeling has predicted a rich landscape of putative extracellular waveforms based on morphology, channel density, and sequential temporal activation. Recent advances in extracellular *in vivo* recording, combining low noise with high spatial density of recording sites, have improved the precision of extracellular waveform measurements, particularly in capturing waveforms beyond the classical somatic spikes, and in some cases, combinations originating from different compartments. This review aims to reorganize extracellular waveform heterogeneity by separating signals stemming from three neuronal compartments using three dimensions: amplitude, duration, and spatial extent or "footprint". In doing so, we argue for a change of perspective, looking beyond somatic spikes to include spatiality and waveform combinations.

# Introduction

All neuronal compartments exhibit specific electrophysiological phenomena that support their specific function. While somata and axon initial segments (AIS) support the initiation of high-amplitude action potentials, axons and dendrites also convey electrophysiological signals of smaller to middle amplitude, sometimes even spikes, on top of pre- and post-synaptic currents that are harder to record in nature. From the first observation of neuronal membrane changes by Bernstein more than 150 years ago (Bernstein, 1868) to the first distinction of different neuronal waveforms in single recordings from stereotrodes (McNaughton et al., 1983), neuronal action potentials have been primarily characterized using single electrodes reporting the electrophysiological signals of a single point of their morphology (usually the soma). Patch-clamp experiments have enabled us to characterize the diversity of electrophysiological signals since the intracellular recordings of action potentials on the giant axon of the squid (Hodgkin and Huxley, 1952).

In the last decades, silicone probes, and particularly high-density extracellular recordings, have become a staple of systems neuroscience, because they yield many neurons simultaneously, they record deep in the brain, which can be done during behavior while maintaining single-spike temporal resolution. This raises the question of whether the diverse repertoire of electrophysiological signals characterized by decades of intracellular recordings can be studied using extracellular electrodes. In other words, whether the shape of extracellular signals can be used to identify the different neuronal compartments that they stem from. In particular, recordings of axonal terminals next to somatic spikes in the same tissue location opens new avenues to study how inputs are locally transformed into outputs, touching there the crux of neural circuit computations.

Recent studies have shown that it is possible to extract other extracellular waveforms shapes using silicone probe (Sun et al., 2021; Someck et al., 2023) or high-density silicone probes (Sibille et al., 2022; 2024), revealing in certain case functional connectivity. Furthermore, high-density silicone probes recording does contain apparently high-quality extracellular waveforms not resembling somatic action potential (Siegle et al., 2021). This prevents the neuroscientific community from utilizing a large portion of the available electrophysiological signals that are seemingly incomprehensible and often set aside as "multi-unit-activity". To address this issue, we propose a simple yet generalizable framework for interpreting the multitude of neuronal waveforms by explicitly considering the three principal neuronal compartments, soma, axon, and dendrites, as the fundamental building blocks for all extracellular signals, particularly when referring to more complex waveforms that are non somatic, likely stemming from combinations, sometimes referred to as "exotic" waveforms.

We aim to compile arguments, theoretical insights, and experimental evidence that demonstrate the feasibility of capturing and measuring extracellular signals from the three ubiquitous neuronal compartments. Through this review, each topic is first introduced from a biophysical perspective and subsequently complemented by experimental evidence, where available. The first part covers the biophysical principles underlying neuronal electrophysiology, linking single-channel activity with membrane waveforms and their corresponding extracellular signals, and is complemented by experimental considerations. The second part then briefly summarizes the modelling and observations of somatic signals, the third part focuses on axons, and the fourth part on dendrites. Finally, in the fifth part, we list some of the known combinations of waveforms from different compartments: dendro-somatic, axon-to-dendrite, and axonal bundles. This outlook should be viewed as an invitation to broaden the search for neuronal waveforms by adopting a compartment-based viewpoint. Considering that the three neuronal compartments activate in a fixed sequence (dendrite to soma to axons, with exceptions such as dendritic backpropagating events), the range of plausible waveform combinations is thereby constrained. Consequently, much of the remaining complexity likely arises from the spatial overlap and interaction of signals originating from these three neuronal compartments.

# Part I: Origins and properties of neuronal electrophysiological signals

## Biophysical principles underlying extracellular signals from neurons

Neuronal electrophysiological signals are mostly based on the $Na^+$ and $K^+$ gradients that are maintained at both sides of the neuronal membranes. Consequently, different currents are mediated by the voltage-gated channels, especially $Na^+$ and $K^+$ which play a predominant role in the generation of action potentials. The waveform produced by these currents can be reproduced with high fidelity using a Hodgkin-Huxley-type (HH) model, which simulates the dynamics of a population of channels using Boltzmann functions, thereby determining their duration and amplitude (Shilnikov et al., 2012). The HH equation describes channel kinetics: The sodium channel opening time constant is on the sub-millisecond timescale, while closing occurs over tens of milliseconds. Potassium time constants are also on the order of a few milliseconds. $Na^+$ and $K^+$ are typically responsible for somatic spikes (Part II). In addition, $Ca^{2+}$ and Cl$^-$ are also involved in neuronal electrophysiological activity, but their channel activations are slow in comparison to those of sodium and potassium. Ca$^{2+}$ can be modeled as a graded response with timescales of 10^-2 to 10^-1 seconds, depending on the channel subtype (Meyer and Stryer, 1991; Foehring et al., 2000; Loewenstein and Sompolinsky, 2003). However, Ca$^{2+}$ currents play a measurable role in the spike waveform during the afterhyperpolarization, especially because of the activation of Ca$^{2+}$-dependent K$^+$ currents (Pineda et al., 1998). Besides, the Ca$^{2+}$ channels with high- and low-voltage-activated subtypes can modulate the firing patterns: the low-voltage-activated subtype can facilitate the sodium channel activation, and high-voltage-activated ones open in response to large membrane depolarization (rev. Simms and Zampioni, 2014). Furthermore, Ca$^{2+}$ channels are heavily involved in the diverse dynamics of dendrites (Part IV), which remain largely inaccessible to extracellular measurements *in vivo*. By contrast, experiments and modeling have both suggested that Cl$^-$ channels have little effect on the neuronal spike shape (Labarca et al., 1985, but see Berndt et al., 2011). Overall, it is worth noting that based on the number, density, and distribution of of Na$^+$ and K$^+$ channels, the resulting extracellular signal can be predicted with biophysics. The channels themselves are spatially organized within the tissue according to the morphology of the neuronal membrane in each compartment.

At the synaptic cleft, the synaptic currents are mediating the transmission of spikes from one neuron to the next. Their dynamics are better described by the ligand-gated channels that depend on one or more conformational changes in the ligand-membrane system. They can be described by quantal- or Markov-chain-type of kinetic models, which quantify the proportion of each possible state of the ligand across the population, while channel permeability can be accounted for using the Goldman-Hodgkin-Katz equation (similar to the Nernst potential but including all permeant ions). This can account for both the kinetics of synaptic current and the possible influence from other neurotransmitters (Destexhe et al., 2003). In particular, the synaptic currents mediated by AMPARs (amino-3-hydroxy-5-methyl-4-isoxazolepropionic acid receptors) and GABA-A-Rs (gamma-aminobutyric acid A receptors) can be fast enough to contribute to the extracellular waveforms, whereas NMDARs-mediated (N-methyl-d-aspartate receptors) currents are typically slower, lasting tens to hundreds of milliseconds (see also Part IV).

From an extracellular perspective, Ohm's law is the theory of choice for predicting the extracellular waveform shape of any neural signal, as the sum of all transmembrane currents from channels in all surrounding compartments, weighted by their distance to the recording electrodes. Transmembrane currents can be related to the biophysical description of the transmembrane voltages as its time derivative (rev. Halnes et al., 2024), and the approximation of the extracellular waveform corresponding to the negative time derivative of the transmembrane voltage has been experimentally validated (Henze et al., 2000). Combining these well-known simplifications leads to the central aspect emphasized here: the systematic incorporation of spatial considerations. In the extracellular space, the time course, amplitude, and spatial arrangement of transmembrane currents will thus define the extracellular waveform. Their summation over space and time is highly linear, independent of whether we choose to use volume-conductor theory or electrodiffusion framework (for an extensive review, see Halnes et al., 2024, Chapter 2). Furthermore, the extracellular medium imposes low-pass filtering on waveforms with increasing distance, while treating all currents similarly (Bedart et al., 2004; Bedart et al., 2022), therefore later spatial considerations from each of the three compartments will hold in the measured extracellular signal. In addition, the amplitude of the waveforms attenuates with distance $r$ according to a $1/r^n$ rule, where the exponent $n$ increases with distance (Anastassiou et al., 2015; Pettersen and Einevoll, 2008; Henze et al., 2000), thereby being related to cell morphology. In short, far from the current sinks and sources (or in their midpoint), the dipole contribution of the electric field is the strongest and its amplitude decays with $1/r^2$. In the vicinity of a current sink or source, there is an additional quadrupole contribution (Halnes et al., 2024), which effectively decreases the exponent $n$. Thus, for somatic waveforms, $1 \leq n \leq 2$ in the vicinity of the soma and $n \geq 2$ far from the soma (Pettersen and Einevoll, 2008). Based on these two principles, we start by inspecting individual neural compartments as quasi-independent contributors of extracellular waveforms, before detailing a few examples of clear summation from two distinct compartments in space and time.

## Experimental evidence linking membrane depolarizations to extracellular waveforms.

The gold standard for membrane waveform measurements at a single point has long been *in vitro* patch-clamp electrophysiology. Patch-clamp recordings can measure single-channel dynamics and currents in an excised patch (Ogden et al., 1994, Chapter 4). They have further confirmed at the single-channel level the validity of biophysical modelling of action potential dynamics and the corresponding extracellular field (Andreozzi et al., 2019; Langthaler et al., 2022). Furthermore, using patch-clamp techniques, decades of studies have monitored the changes in membrane potential at the soma, particularly the generation of action potentials at the AIS, and its propagation along the axon.

There is a justified simplification in interpreting these single-point measures of neuronal activation as reflecting the co-polarization of the neighboring neuronal morphology. Experimentally, it is referred to as the "space clamp", i.e., the assumption that the axial currents in the membrane can be neglected across some distance. Consequently, patching the cell at two different points goes with the assumption of equal changes of potentials in both locations. Such a co-activation has been repeatedly confirmed by patching multiple points of the same neuron (Buccino et al., 2024; Larkum et al., 1999; Sasaki et al., 2021). Theoretically, the passive decay of the membrane potential is characterized by the space constant $\lambda$, which is inversely related to the dimensionless electrotonic length of cables. Biophysical modelling has established that the space constant of the passive neuronal cable ranges from 100 to 500 µm (Rall, 1969), and that it shows some frequency-dependence (Bedard and Destexhe, 2008, Ilmoniemi et al., 2016). With active currents in the membrane, the space constant of a neuron typically becomes shorter due to the reduction in membrane resistance upon the activation of ion channels. We advocate here for looking beyond this space clamp, i.e., to different neuronal compartments giving rise to distinct extracellular waveforms, as high-density silicone probes permit such distinction. In particular, Neuropixels probes can sample the extracellular signal at 30 kHz along several millimeters, thereby monitoring different points well beyond the neuronal space constant and enabling the monitoring of neuronal activation on two opposite sides of the same neuron. Altogether, patch-clamp techniques have established the nature, dynamics, and properties of membrane waveforms, which we hereafter distinguish explicitly from extracellular waveforms.

In the late 90s and early 2000s, tetrodes and silicon probes became the gold standard for extracellular measurements of neuronal activity *in vivo*, as they could reliably extract somatic action potentials. Unlike single-point measurement, tetrodes and silicon probes record extracellular activity from several spatially nearby neighboring channels, measuring extracellular waveform at an estimated distance of about 150 μm between the soma and recording sites (Harris et al., 2000; Henze et al., 2000). This feature enables us to distinguish waveforms from different somata based on the shape and amplitude of the action potential in each channel by triangulating different action potential sources (rev. Buzsaki, 2004). Interestingly, tetrodes and silicon probes have mostly reported a single waveform shape per neuron, based on the assumption that most extracellular signals originate from the somatic action potential (but see, Someck et al., 2023; Senzai et al., 2019; Sun et al., 2021; Bereshpolova et al., 2007). In addition, probe impedance has only a small effect on spike amplitude (Nelson and Pouget, 2010; Neto et al., 2018), but it significantly affects the signal-to-noise ratio of the recording (Berntd et al., 2011; Neto et al., 2018). Currently, high-density silicon probe techniques use impedance adaptations, thereby improving the signal-to-nose ratio and lowering the threshold for extracellular detection (Jun et al., 2017; Steinmetz et al., 2021; Ye et al., 2025). We claim that the improved resolution offered by high-density silicon probes will permit an increase in the acuity of available measurements of neuronal signals. In particular, these probes capture signals approximately every ten micrometers over several millimeters. Therefore, this high spatial density enables the identification of extracellular signals from distinct points of the neuron that are virtually independent, particularly at points located at distances that exceed the neuronal membrane's space constant. Consequently, the waveforms from three different neuronal compartments *in vivo* are being distinguished, highlighting the new technical possibility to measure them independently. Finally, biophysical considerations can also predict the existence of extracellular signals with amplitudes that may currently lie below technological or methodological detection limits.

# Part II Waveforms from the soma and proximal AIS

## Biophysics of somatic action potentials

Modeling has been a useful approach in guiding research directions and providing insights into experimental observations. For example, Mainen et al. (1995) successfully predicted that the sodium channel density in the axon initial segment (AIS) needs to be noticeably higher than that in other compartments (dendrites and soma) for the action potential to be initiated at the AIS. Despite some initial contradictory experimental observations (Colbert and Johnston, 1996), the prediction was confirmed with other experimental approaches: The AIS has up to 50-fold higher sodium channel density than other compartments (proximal dendrites and soma; Mainen et al., 1995; Kole et al., 2008), together with faster kinetics of the sodium channels (Colbert and Pan, 2002), and consequently the largest spike amplitude (Kole et al., 2008; Hu et al., 2009). The typical extracellular spike of a neuron is traditionally thought to originate from somatic depolarization, which can be difficult to distinguish from that of the AIS, given their usual spatial proximity. Furthermore, modeling has predicted that the AIS is the driving force for the soma-AIS dipole that manifests as a spike in the extracellular voltage (Teleńczuk et al., 2018), as confirmed experimentally *in vitro* (Bakkum et al., 2019). Therefore, we do not distinguish between somatic and AIS spikes, instead we consider them as originating from the same compartment and refer to them as somatic spikes, unless required by the context. In addition, depending on the neuronal type and associated genetic expression pattern, biophysical models can account for the differences in neuronal firing patterns (Schreiber et al., 2004; Reva et al., 2025) and waveforms of the membrane voltage (Gouwens et al., 2019; Beau et al., 2025). Computational studies have further proposed that the action potential waveform can vary with morphological variability within a single neuronal type (Sandbote et al., 2025).

As stated previously, the extracellular voltage spike is the negative time derivative of the transmembrane voltage (Anastassiou et al., 2015), where the extracellular spike's negative component stems from fast inward $Na^+$ and outward $K^+$ currents, and the subsequent positive component stems from the slower outward $K^+$ currents, giving rise to the temporally asymmetric waveform shape (Fig 1). The spike amplitude can be approximated as decreasing inversely with distance from the soma (Anastassiou et al., 2015; Harris et al., 2000; Henze et al., 2004), with potential effects from the dendritic morphology and the dendrite-to-soma cross-sectional area (Pettersen and Einevoll, 2008). In general, extracellular signals experience a low-pass filtering effect, where higher frequency signals attenuate more strongly in comparison to the low-frequency signals (Pettersen and Einevoll 2008; Leski et al., 2013). For more details on this extracellular filtering of the membrane's waveforms, please refer to the previous Part I.

## Experimental measures of somatic spikes

Spike measurements *in vitro* have ranged from primary neuronal cultures to more acute preparations, or organotypic slices. Using either patch clamp pipettes, Multielectrode arrays or high-density CMOS (complementary metal-oxide semiconductor), neural activity has been routinely measured in many neuroscience laboratories, contributing to a vast amount of neuroscientific knowledge. Such measurements have recently been extended using high-density multielectrode arrays combined with multiple patch clamps on different neural points to confirm the validity of such approaches (Buccino et al., 2024).

Spike measures *in vivo* were initially performed with juxtacellular recordings (Dowben and Rose, 1953), enabling biocytin staining of the recorded neuron (Pinault, 1995). Since then, advances with stereotrodes, tetrodes and the many available different silicon probes have enhanced the possibilities to measure neuronal action potential in living tissue of anesthetized and awake animals (Neuronexus ref.; rev. Buzsaki, 2004). Noticeable achievements have been reached in few instances where a single neuronal juxtacellular recording is associated with neighboring silicon probe measures (Harris et al., 2000). Consequently, the variety of spike shapes has been utilized heavily in neuroscience to distinguish different neuron types. In particular, different neuronal waveform shapes have been related molecularly to different neuronal classes (Gouwens et al., 2019; Beau et al., 2025), including inhibitory neurons (rev. Klausberger et al., 2008; Tzilivaki et al., 2023), which are of particular interest because of their ability to strongly influence local networks in the cortex (Bugeon et al., 2022) and the hippocampus (Valero et al., 2025). Single-spike waveforms were early on distinguished between narrow, fast spiking waveforms or broader waveforms (Mountcastle et al., 1969), as later confirmed in guinea pigs (McCormick et al., 1985), cats (Azouz et al., 1997), and rodents (Simon et al., 1978). Such a distinction is still used today as an approximation for experimentally distinguishing some fast-spiking inhibitory neurons from other neurons (Senzai et al., 2019), using the extracellularly measured peak-to-trough duration. More parameters characterizing cell types based on extracellular spike waveforms have been refined in more extensive studies of extracellular waveform parameters (Lee et al., 2021; Haynes et al., 2024).

Further results have confirmed a partial overlap between narrow waveforms and some inhibitory neuron types (Valero et al., 2025). A wealth of methods has been developed over the last two decades to separate different neuron types based on waveform shapes, which is not covered in this review. These methods are closely related to spike sorting, as detailed in the next sections, and include more automated classifiers based on artificial neural networks (Beau et al., 2025).

## Spike sorting

Spike sorting is an approach that distinguishes single extracellular somatic waveforms originating from different neurons that were recorded simultaneously. It was first proposed and used by Gerstein and Clark (1964) to distinguish the contribution of the action potential waveforms of several neurons in the signals recorded with a tungsten microelectrode. Since then, multiple approaches have been developed to improve clustering quality and efficiency, keeping pace with advances in hardware technologies that enable the use of hundreds or even thousands of electrodes. When using multiple probes, this enables the recording of up to 800 units simultaneously across multiple brain regions (Siegle et al., 2021). Among the popular algorithms are Kilosort (Pachitariu et al., 2024), SpyKING CIRCUS (Yger et al., 2018), HDSort (Diggelmann et al., 2018), IronClust (Zhang et al., 2022), Mountain sort (Chung et al., 2017), HerdingSpikes (Hilgen et al., 2017), WaveClus (Quiroga et al., 2004), KlustaKwik, MClus, and OSort (Rutishauser et al., 2006).

In conclusion to Part II, we simplify somatic spike waveforms that are expected to originate from the AIS and have an asymmetric shape, starting with a faster sodium-driven depolarization followed by a longer potassium-mediated repolarization (Fig. 1). Different cell types can show their own waveforms, as often used for spike sorting, but the typically large extracellular amplitude clearly differentiates the somatic spikes from the other compartments, as detailed in the next parts (Fig. 4).

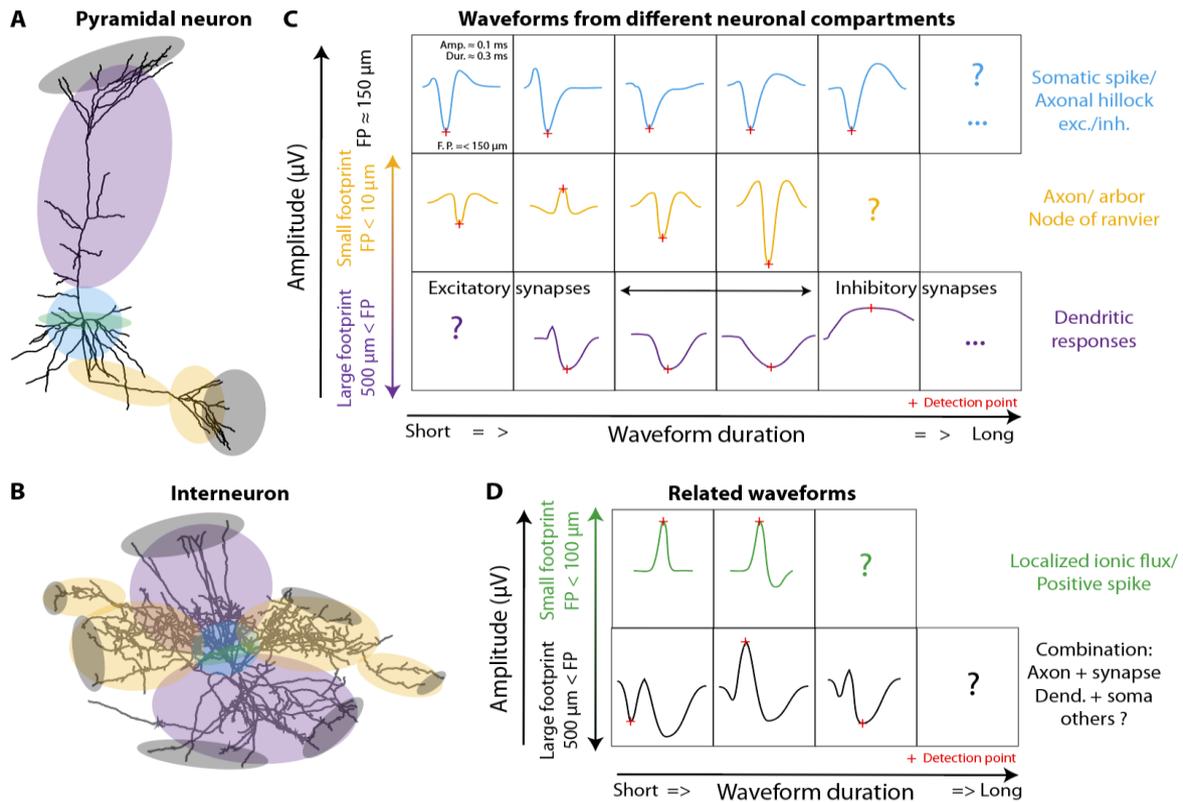

*Figure 1: Schematics of extracellular waveforms from the three neuronal compartments:* *A simplified schematic illustrating three distinct neuronal compartments and their corresponding different schematic waveforms.* ***A****. Morphology of an excitatory pyramidal neuron and the location of its different compartments.* ***B****. Morphology of a Parvalbumin-positive GABAergic interneuron and the location of its different compartments (reconstruction adapted from Tzilivaki et al., 2019, Tzilivaki et al., 2025).* ***C****. Simplified organization of waveforms from the three different neuronal compartments: somatic (top, blue), axonal (middle, yellow), and dendritic (bottom, purple) origins.* ***D****. Limited overview of neuronal compartment-related waveforms measured extracellularly, stemming either from spiking-related currents (top, green) or from combinations of axonal arbor and synaptically evoked dendritic extracellular waveforms.*

# Part III Axonal waveforms

The axon can consist of four distinct sub compartments that should be distinguished: single axonal projections, nodes of Ranvier (if myelinated), axonal arbors (if present), and presynaptic boutons. Different waveforms are associated with these four sub compartments, which can be distinguished. However, in experiments, it is unlikely that these waveforms would overlap in a single extracellular recording site. For this reason, we treat each axonal sub compartment separately, first from a biophysical perspective and then through experimental evidence.

## Axonal projections

The intracellular mechanics of axons were first modeled in the seminal work of Hodgkin and Huxley (1952), which has profoundly influenced subsequent modeling of all neural compartments. Cable theory (Rall, 1962; 1969; 1977), although initially developed for dendrites, soon offered a much simpler analytical framework for studying axonal extracellular fields. The tri-phasic symmetry of the axonal waveforms was identified early (Clark and Plonsey, 1968; Plonsey, 1969; Stegeman, 1979). Yet, biophysical predictions have detailed that any perturbation in the symmetry of the axonal morphology inevitably leads to small asymmetries in the corresponding extracellular waveform (Clark and Plonsey, 1968; Stegeman et al., 1979; 1997). Such perturbations include (among others, Stegeman et al., 1997): directional changes of the axon, local conductivity variations in the myelin or in the extracellular space, changes in the internodal distances in myelinated axons, myelination differences in the paranodal and juxtaparanodal region (Chiu and Ritchie, 1981; Waxman and Ritchie, 1985), and axonal branching (McColgan et al., 2017). The amplitude of a single axon's extracellular waveform was modeled to be small — on order of tens of microvolts — compared to other neuronal compartments (Clark and Plonsey, 1968). A straight axon, approximated as a line source, generates primarily a quadrupolar electric field (Plonsey and Barr, 2007, Chapter 8) whose amplitude decays rapidly with distance. Thus, axons have traditionally been considered negligible contributors to extracellular signals. Yet, axonal arborizations and synchronously activated axon bundles can sum their extracellular contribution, which is roughly scaling with the number of branches or axons involved (Fig. 2), thus often producing waveforms with amplitudes up to the millivolt range and leading to "axonal volleys" in bundles (Part V, Section 3).

Experimentally, the capability to measure and isolate activity from single axonal projections has been facilitated *in vitro* by two main factors: sparse neural density, which leads to sparse presence and makes the identification of single axons possible (Li et al., 2015). Consequently, *in vitro* recordings of axonal activity over recent decades have provided detailed insights, for example in retinal physiology (Zeck et al., 2011). In the hippocampus, double patch recordings of the soma and an axonal projection (Sasaki et al., 2021) revealed nearly symmetric short-duration axonal waveforms (0.05 - 0.15 ms). *In vivo*, symmetric short-duration axonal waveforms have been repeatedly observed in so-called "en passant axons", either exhibiting a classical negative waveform (Someck et al., 2023; Beau et al., 2025; Sibille et al., 2022; Ye et al., 2025) or displaying a positive peak with an amplitude exceeding 50 µV (rev. Robbins et al., 2013; Barry, 2015; Someck et al., 2023; Sun et al., 2021; Senzai et al., 2019). These positive waveforms have been also associated with return currents around the pyramidal layer (Someck et al., 2023). There is further evidence for extracellular detection of "en passant axons" in the rat thalamus (Bartho, 2021), the avian thalamic equivalent (Goldberg and Fee, 2011), the avian auditory nerve (Kuokkanen et al., 2025), the rat entorhinal cortex (Robbins et al., 2013), the cat visual cortex (Sun et al., 2021), the cerebellum (Khaliq and Raman 2005; Beau et al., 2025), the visual cortex (Senzai et al., 2017), the optic tract (Schröder et al., 2020), and the mouse hippocampus (Someck et al., 2023). High-density multielectrode arrays have further enabled the tracking of spike propagation along axonal projections, either anterogradely (Li et al., 2015) or antidromically (Su et al., 2024), strengthening the approximation that extracellular axonal projection waveforms are symmetric. It is not to be undermined that slightly asymmetrical axonal waveforms have also been long reported experimentally (Galambos and Davies, 1943). An additional key analytical feature used in some studies is the time delay between the somatic and axonal spikes, captured in the inter-waveforms cross correlogram (CCG) computed from recordings sites located on two different shanks of the same silicon probe (Someck et al., 2023; Senzai et al., 2019). Tangential Neuropixels insertions in the superior colliculus (Sibille et al., 2022) and the cortex (Sibille et al., 2024) of mice have revealed axonal projections that precede the detection of axonal arbors, with waveforms characterized by short durations and pseudo-symmetric shapes, consistent with direct retinal observations in the optic track (Schröder et al., 2020). However, a broader diversity of axonal projection waveform has been reported (Schröder et al., 2020; Ye et al., 2025), reflecting possible variability in morphology and recording placement around the axons, as observed by other authors (Sun et al., 2021; Lee et al., 2021; 2023; Senzai et al., 2019).

## Axonal nodes of Ranvier

In myelinated axons, nodes of Ranvier are of particular interest due to their very high concentration of ion channels. The local clustering of $Na^+$ and $K^+$ channels is essential for the function of the node of Ranvier (Nelson and Jenkins, 2017), exhibiting species-dependent variability of kinetics (Marks and Loeb, 1976). In practice, spikes propagating along myelinated axons can be detected extracellularly only when the recording electrode is right in the vicinity of a node of Ranvier. When low-impedance electrodes capture currents from a large spatial volume, the effective average channel density is low, resulting in a small extracellular amplitude from myelinated axons (Waxman and Ritchie, 1985; Waxman and Ritchie, 1993; rev. Tanner and Tzingounis, 2022). This effect is enhanced by the line-source geometry of extended axons, which primarily produces quadrupole fields. However, high-impedance electrodes with a small recording radius will capture the very high local density of $Na^+$ and $K^+$ channels at the node of Ranvier (Ritchie and Rogart, 1977; Amor et al., 2017), generating an extracellular waveform that is spatially restricted but substantially large in amplitude (Tasaki, 1964; Lopreore et al., 2008; Radivojevic et al., 2017). This principle was recently exemplified by the newly released Neuropixels Ultra probe, which has recording sites spaced 6 μm apart (Ye et al., 2025). Using this ultra-dense recording technique, the authors reported a novel class of extracellular waveforms resembling classical spikes, but with the combined characteristic of both high amplitude (>50 μV) and a very small spatial footprint (less than 20 μm in size), along with symmetrical waveform shapes. Pharmacological controls suggested that these signals originate from axons, thereby leading to a working hypothesis for the extracellular signature of nodes of Ranvier (Ye et al., 2025). Such small-footprint, large-amplitude waveforms have also been observed across multiple vertebrate species and brain regions (Ye et al., 2025).

In summary, larger amplitude waveforms, associated with smaller spatial footprints, may serve as a distinguishing signature for nodes of Ranvier — if further confirmed — paving the way for more systematic spatial consideration to distinguish different neuronal compartments.

## Axonal Arbors

As action potentials reach the end of the axonal projections, they often encounter complex and sometimes dense axonal arbors. As discussed above, idealized line sources produce nearly symmetric extracellular waveforms (Clark and Plonsey, 1968; Plonsey, 1969; Stegeman, 1979). However, any symmetry-breaking of the axonal anatomy or morphology leads to a symmetry break of the waveform, although these waveforms remain relatively symmetric compared to those from other cell compartments. Consequently, our current simplification states that all axonal waveforms are short and nearly symmetric (Fig. 4). Most importantly, axonal arborization effectively multiplies transmembrane currents by the number of branches carrying simultaneous currents (McColgan et al., 2017). Possible changes in myelination, channel densities, and axonal diameters can gradually modulate the extracellular waveforms across the entire arborization (Tomassy et al., 2014; Fulcher et al., 2019; Liu et al., 2024), leaving the multiplication principle unaffected. The density and spatial orientation of the arborization strongly influence near-field extracellular waveforms. On one hand, it can produce middle-amplitude signals, possibly with tri-phasic waveforms, when facing highly parallel branching patterns (McColgan et al., 2017). On the other hand, it can produce small-amplitude, short-lived bi-phasic extracellular waveforms when facing a branching pattern with mixed (highly asymmetric) spatial orientations (Radivojevic et al., 2017; Ye et al., 2025, Fig. S15). Additionally, low-frequency components may emerge from the interplay between the arborization size and the conduction velocity of the axon (McColgan et al., 2017), although these components are primarily observable in the far field and thus irrelevant for spike detection.

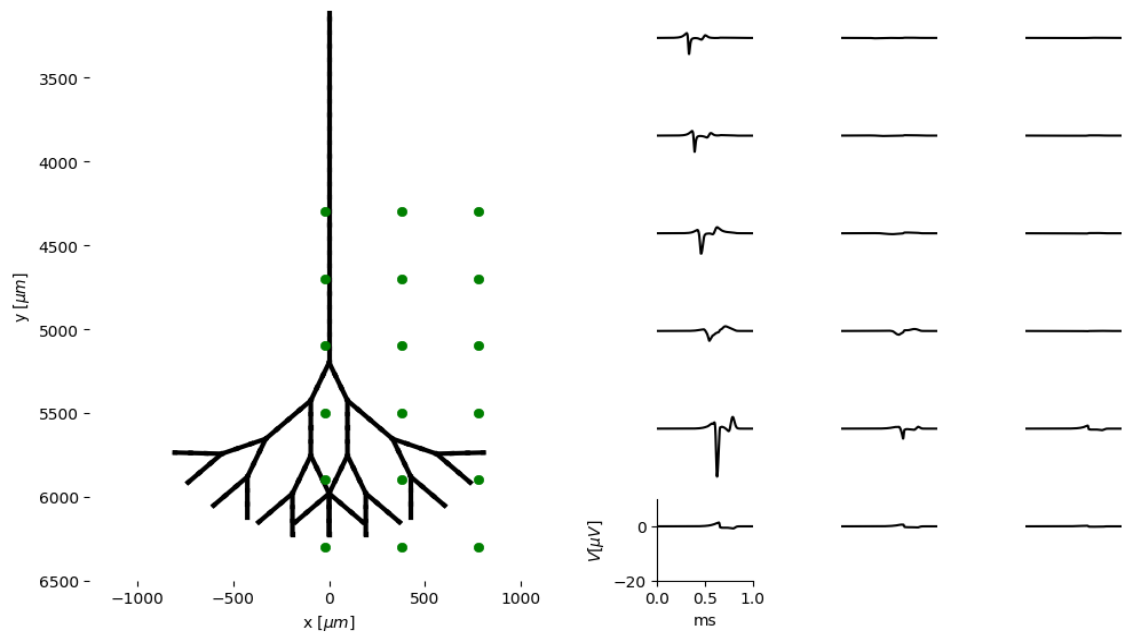

*Figure 2: Attenuation of the extracellular electrophysiological signal over distance:* Effects of recording distance and source density on the waveform from the axonal arborization. NEURON / LFPy modeling of a branching myelinated 2D axon with active $Na^+$ conductance in the nodes of Ranvier: EFP at the indicated recording locations in a homogeneous medium. Inter-electrode interval in both directions: 400 um, closest column to the axon projection: x = -20 um. Methods: code as published in McColgan et al., (2017, https://github.com/phreeza/pyLaminaris, MIT License), with following parameters changed for a slightly modified axon morphology: axons.mtype = 'bif_term', np.random.seed = 1, axons.dir_mean = 174, axons.ang_var = 20, axons.depth = 4.

Experimentally, axonal action potentials followed by postsynaptic responses have been detected on long-range axons in turtle brains *ex vivo* (Shein-Idelson et al., 2017). Excitatory and inhibitory axonal projections are followed by characteristic negative excitatory or positive inhibitory synaptic responses, respectively, with the first symmetric peak reflecting the axonal action potential (Shein-Idelson et al., 2017). Using Neuropixels *in vivo*, direct detection of axonal arbor have been realized in the superior colliculus and visual cortex of mice, aided by the favorable planar morphology of the axonal projection and the tangential alignment of the inserted probe (Sibille et al., 2022; Gehr et al., 2023; Sibille et al., 2024). Spike-triggered local field potential (LFP) analysis (multichannel waveforms) further revealed axonal projections adjacent to the probe within the superior colliculus, immediately prior to the axonal arbor detection (Sibille et al., 2022; Gehr et al., 2023). The axonal arbor's action potential is followed by the consequent postsynaptic response in the superior colliculus and cortex (Sibille et al., 2022; Gehr et al., 2023; Sibille et al., 2024). In these studies, axonal waveforms were symmetric, negative, and of moderate amplitude (<75 µV in the superior colliculus, <50 µV in the mouse cortex, Fig. 3), consistent with recordings from unmyelinated axonal segments.

The visual system provides a particularly informative model to study sensory pathways, due to its simple anatomy and accessibility. During development, the visual pathway undergoes three stages of spontaneous retinal waves (rev. Huberman et al., 2008; Xu et al., 2016; Ackman et al., 2012; Gribizis et al., 2019) that progressively reshape and refine the axonal arbors (Ackman et al., 2012; Chandrasekaran et al., 2005; Chandrasekaran et al., 2007). In addition, cortical activity becomes decoupled from the thalamic inputs during stage III retinal waves (Gribizis et al., 2019). It has been confirmed that thalamocortical axonal arbor spreads are relatively large (Antonini et al., 1998, 250-500 µm), as confirmed by measuring the spatial footprint in Neuropixels recordings (450 µm, Sibille et al., 2024). In contrast, retinotectal axonal arbors exhibit a middle-range footprint when recorded with high-density silicone probes (250 µm, Sibille et al., 2022). Given the growing wealth of large-scale electron microscopy (EM) reconstructions of axons and dendrites (Ding et al., 2025; Weis et al., 2025), it may soon become feasible to systematically correlate known axonal arbor anatomies (Mazade and Alonso, 2017) with experimentally observed extracellular waveforms across pathways such as thalamocortical projections (Swadlow and Gusev, 2000; Bereshpolova et al., 2019; Su et al., 2021; Sibille et al., 2024), retinotectal pathway (Sauvé et al., 1995; Sibille et al., 2022; Gehr et al., 2023), and corticotectal projections (Su et al., 2024). These observations highlight the need for new biophysical models capable of predicting weaker signals from smaller or less complex axonal arbors.

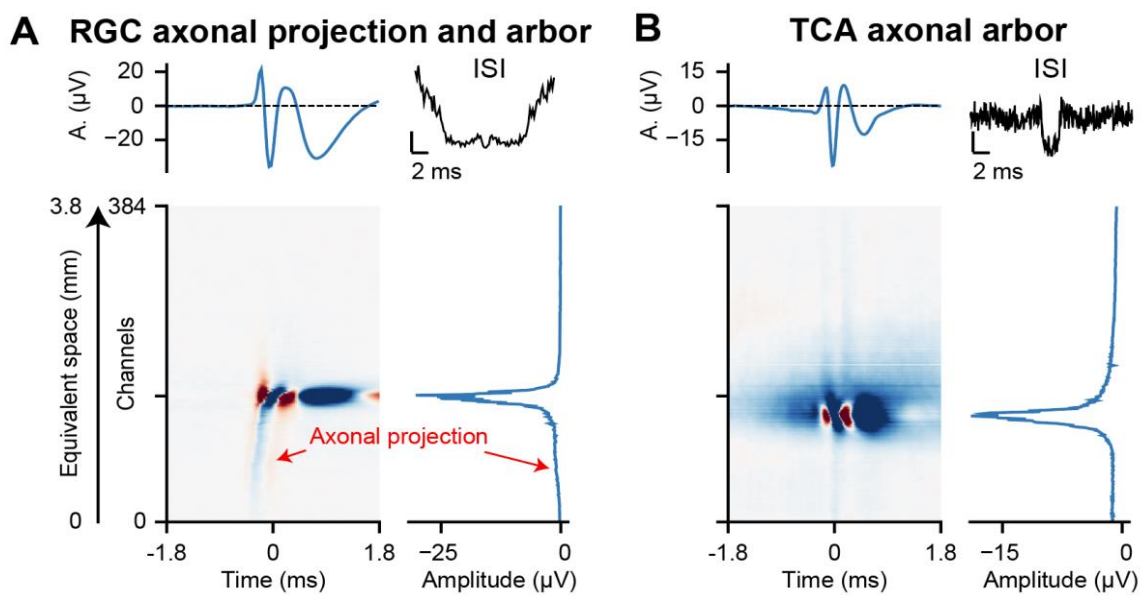

*Figure 3: Measurements from an axonal arbor and from a pre-activated axonal projection:* *Axonal projection and axonal arbor symmetric waveform are middle in amplitude but large in space. Experimental details as in Sibille et al. (2022) for retinal ganglion cells (RGC) and in Sibille et al. (2024) for thalamocortical axons (TCA).* ***A****. Classical single-channel waveform representation (top left) and the corresponding Inter spike interval (ISI, top right). Classical multi-channel waveform obtained from averaging the raw signal for all spikes illustrates the long extent of the axonal projection ~1 mm (bottom left). Corresponding amplitude along the probe shows that axonal projection in this probe position and alignment is below 5 µV (bottom right).* ***B****, Similar plotting for a TCA. All results shown are under the* [CC-BY-4.0 license](CC-BY-4.0 license)*.*

In summary, axonal arbors often, but not exclusively exhibit near-symmetric waveforms with short durations (<0.2 ms) and middle to large amplitudes compared to axonal projections, as each branch additionally produces extra currents. The axonal arbor extracellular waveforms are often followed by postsynaptic responses, when spatially adjacent to the electrode (Part V, Section 2). We expect that the larger morphology of the axonal arbor will be reflected in a larger extracellular footprint due to the spatial summation of synchronous signals, allowing detection either in the vicinity or at a close distance (Fig. 5).

## Presynaptic Terminals

Once an action potential reaches the end of its axonal projection, presynaptic boutons will be almost concomitantly activated. Consequently, if the spatial density of axonal presynaptic boutons is sufficiently high, a signal should be detectable. In practice, it is difficult to disentangle the extracellular signals arising from the axonal arbors from those produced by near-synchronous presynaptic bouton currents, due to their close spatial proximity. Thus, experimental evidence of extracellular waveform(s) recorded exclusively from presynaptic boutons —and not from the axonal arbor — is very rare. A notable exception is the calyx of Held synapses in the auditory brainstem, where presynaptic contacts of the single giant synapse are spatially very dense, surrounding the postsynaptic cell body with hundreds of synaptic contacts that are activated simultaneously. Upon the arrival of the axonal spike, a large extracellular signal, so-called prepotential, can be observed. Prepotentials are typically short (few hundred microseconds), monophasic in waveform (Pfeiffer, 1966; Zhang and Trussell, 1994; Kopp-Scheinpflug et al., 2003; Englitz et al., 2009; Kuokkanen et al., 2025), and reliably followed by the somatic spike within a typical synaptic delay (about 50-100 µs). Their polarity can either be the same as the postsynaptic somatic spike or be reversed. Prepotential extracellular amplitude can be as high as that of the somatic spike, depending on the relative position of the recording electrode with respect to the AIS of the postsynaptic cell body (McLaughlin et al., 2008; Kuokkanen et al., 2025). Overall, it is important to verify that prepotentials are to be distinguished from postsynaptic current, which can produce extracellular waveforms of larger amplitudes but longer durations (cf. below). These considerations are shedding more light on our central premise: extracellular waveforms can originate from different neuronal compartments, in particular their possible combinations.

In conclusion of part III, axonal waveforms are near-symmetric, predominantly negative across axonal projections, nodes of Ranvier, and axonal arbors. Nodes of Ranvier may exhibit middle- to large-range amplitudes (>50 µV) and a small spatial footprint (< 20 µm, Ye et al., 2025), whereas axonal arbors' extracellular waveforms tend to have small- to middle-range amplitudes (20-75 µV) but large (250 µm in the superior colliculus, Sibille et al., 2022) to very large spatial footprint (450 µm in the cortex, Sibille et al., 2024). This is to say, the signals can presumably be recorded within the arborization ("synchronization radius", Kuokkanen et al., 2010; Lindén et al., 2011), but not within an arbitrary radius of the size of the footprint. This contrast illustrates one of the key principles emphasized here: spatial characteristics provide critical leverage for distinguishing extracellular waveforms arising from different axonal sub-compartments (Fig. 4).

# Part IV Dendrites

Dendrites constitute the main input sources of excitatory and inhibitory neurons and they exhibit striking complexity in their morphology. Dendritic trees extend up to a millimeter long, host of synaptic boutons, and are responsible for integrating incoming excitatory and inhibitory synaptic inputs from many different axonal terminals, as recently reconstructed from EM in mouse V1 (Weis et al., 2025). This section aims to explore the multifaceted mechanisms by which dendritic morphological features, in conjunction with active membrane properties, shape the electrophysiological signature of neurons and therefore contribute to the diversity and intricacy of recorded extracellular waveforms. Dendritic activation is typically slower than classical action potentials originating from the axon initial segment. This temporal slowness renders their direct extracellular detection almost impossible with current methods, as most spike sorting algorithms screen for events shorter than a few milliseconds (cf. part II, Pachitariu et al. 2024). These reasons are likely underlying the several converging publications proposing that, from a biophysical perspective, dendritic processes are the dominant contributors to the LFP signals from pyramidal cells (Sinha et al., 2022; Ness et al., 2016, Liao et al., 2024) but also from interneurons, based on recent computational predictions (Tzilivaki et al., 2025). Here we propose a reduced outlook of how current knowledge in biophysics and neuronal physiology may guide toward future identifications of single dendritic events. In particular, we ask whether known dendritic activities could, in principle, be directly detected extracellularly? Combined small amplitude and longer durations of dendritic signals appears to hinder this possibility, with present-day technology. Furthermore, we suggest that single dendritic activations may already contribute in most electrophysiological recordings, under the indirect form of combined waveforms, that we refer to as "exotic waveforms". We then further discuss whether newer detection methods could be tailored specifically to identify extracellular dendritic responses. This part is organized in four sections ordered by increasing dendritic temporal duration: fast sodium-dependent dendritic spikes, backpropagating action potentials (bAPs), synaptically evoked responses, and longer lasting dendritic nonlinear responses, e.g., calcium spikes and plateau potentials.

## Dendritic sodium spikes

The electrical behavior of dendrites is substantially augmented by a diverse array of voltage-gated ion channels and ligand-gated receptors (rev. Lai and Jan, 2006; Sjöström et al., 2008), whose densities and spatial distributions vary across the dendritic trees and among different neuronal types. These active properties effectively transform dendrites from passive recipients of synaptic input into dynamic and highly nonlinear computational elements endowed with the capacity to modulate, amplify, and, in certain instances, initiate electrical signals (Poirazi et al., 2003; Larkum et al., 2001; Tzilivaki et al., 2019; Tzilivaki et al., 2022; Stuart and Spruston, 2015). Consequently, under certain circumstances dendrites are capable of generating localized regenerative electrical events, so called dendritic sodium spikes. For example, numerous ion channel subtypes contribute to dendritic excitability as for $Na^+$ channels which are expressed in the dendrites of many neuron types. They can facilitate the initiation of localized dendritic $Na^+$ spikes, particularly within thicker dendritic branches or at "hotspots" of high channel density (Williams and Stuart, 2000; Castanares et al., 2020).

These are relatively fast spikes mediated by voltage-gated Na+ channels and will look like more classical spikes (rev. Stuart and Spruston, 2015), albeit having slightly longer duration compared to somatic and axonal spikes (Destexhe et al., 2022). They can be initiated in thicker dendrites and may serve to amplify distal inputs or to facilitate signal propagation across branch points. These short lasting dendritic phenomena (<2 ms) have been broadly reported *in vitro,* initially in the cerebellum of alligator (Llinas et al., 1968), in the dendrite of hippocampal neurons of rat (Kamondi et al., 1998; Bittner et al., 2017 — in combination with $Ca^{2+}$ spikes), and then in avian Purkinje cells (Llinas et al., 1976) and globus palladium of rats (Hanson et al., 2004). Their confirmation *in vivo* has been more limited to the context of longer lasting dendritic spikes where sodium dendritic spikes may feed dendritic spikes (cf. next sections). Because these small-amplitude action potentials are commonly observed intracellularly, their detectability in the extracellular space remains an open question. Given the low transmission from the dendrite and the soma (Larkum et al., 2007), it is more plausible that dendritic sodium spikes primarily contribute to longer-lasting dynamics of dendritic dynamics such as plateau potentials, described in the next sections.

## Backpropagating action potentials

From a biophysical perspective, the smooth coverage of Na+ and K+ channels along the dendrites implies that somatic action potentials can partially invade the dendritic tree, producing bAPs. The bAPs are strongly modulated by dendritic morphology (e.g., branch diameter, tapering ratio) and by the expression of voltage-gated ion channels (Williams and Stuart, 2000). For example, it is known that A-type channels are often more densely expressed in distal dendrites (rev. Jerng et al., 2004) which can substantially attenuate bAPs amplitude (Hoffman et al., 1997).

A number of *in vivo* experimental insights of bAPs emerged from silicon-probe recordings aligned vertically along the dendritic tree of cortical pyramidal neurons (Buzsaki and Kandel, 1998; Bereshpolova et al., 2007), which was later detailed using high-density silicon probe (Jia et al., 2019). These authors zoomed the signal around the soma to observe the dendritic depolarization following the somatic spike, reporting an extracellular waveform resembling that of the somatic spike. They characterized a backpropagation in the dendrite which exhibits a drastic amplitude decay with at least 80% attenuation in amplitude 300 μm away from the soma (Bereshpolova et al., 2007), when not reporting a stronger decay (Jia et al., 2019). Because bAPs have the same extracellular waveforms as somatic spikes, their direct detection, independent of somatic spikes, may be very challenging. Nevertheless, bAPs could rather be systematically used to assess whether a somatic extracellular waveform has any "dissipative side", which would indicate the location of the neuronal dendrite (Fig. 4F).

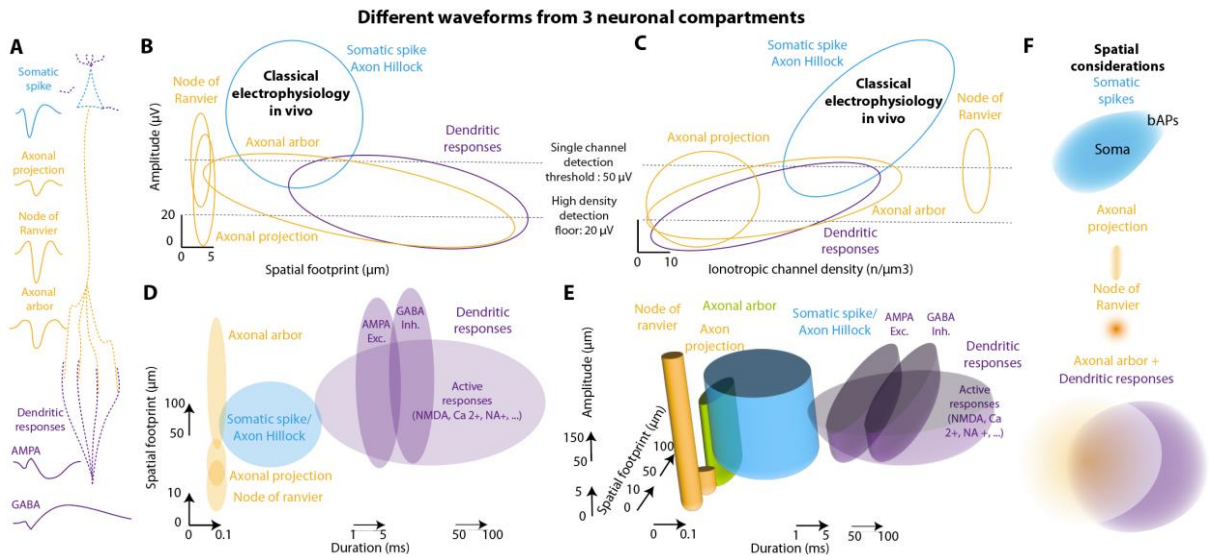

*Figure 4: Biophysical prediction of the three waveform amplitudes and spatial footprints:* Three distinct neuronal compartments producing many different waveforms. **A**, Schematics of different neuronal compartments from the soma (blue), axonal projections, nodes of Ranvier, axonal arbors (yellow), and dendritic responses (purple). **B**, Representation of the expected relationship between amplitude and the morphological spreads of the different neuronal compartments. **C**, Representation of the expected relationship between the amplitudes against the density of the ionotropic channels in the different neuronal compartments. However, see also Fig 2 for an example of the variability of waveforms due to spatial summation. **D**, The scheme illustrates how spatial footprints and durations may separate certain features of extracellular waveforms. **E**, 3D representation schematizing the relationship between spatial footprint, amplitude and durations of the extracellular waveforms. **F**, Expected spatial decay of the waveforms from the three neuronal compartments.

## Synaptically-evoked dendritic responses

Apart from voltage-gated ion channels, dendritic electrophysiological responses are strongly modulated by ligand-gated ion channels, including AMPARs, NMDARs, and GABARs receptors. AMPARs mediate fast excitatory postsynaptic currents exhibiting rapid activation time (0.2-0.8 ms), and deactivation time (1.3-2.0 ms, Kleppe et al., 1999). In contrast, GABARs mediate fast inhibitory currents, leaving $Cl^-$ ions in which hyperpolarize the post-synaptic dendrite. GABARs have a fast activation time like AMPA receptor, but longer de-activation time constants, ranging from 20 to 600 ms depending on the receptor isoform (rev. Sallard et al., 2021). Modelling predicts that AMPA and GABA-synaptically evoked currents reach an amplitude of tens to hundreds of picoamperes, resulting in extracellular fields in the order of 5-10 μV (Hagen et al., 2017), which place them theoretically at the limit of extracellular detectability due to inherent neural and thermal noise *in vivo*. Both of these currents passively propagate to the soma while undergoing a distance-dependent attenuation governed by dendritic cable properties (Bédard et al., 2008). High axial resistance of fine dendritic segments combined to the inherent low pass filtering characteristics of the neuronal membrane will together facilitate semi-autonomous computational operations in distinct dendritic subregions (rev. Reyes, 2001; Stuart and Spruston, 2015). Consequently, forward propagation from distal inputs to the soma is undergoing considerable attenuation unless counteracted by the presence of active conductance (rev. Reyes, 2001; Remme and Rinzel, 2011).

Experimentally, the various neurotransmitters (AMPA / NMDA / glutamate, GABA / glycine / serotonin) and their corresponding receptor subtypes exhibit different kinetics which have been related to different LFP kinetics (rev. Valbuena and Lerma, 2016), which therefore can underlie different functions (rev. Womelsdorf et al., 2014). In particular, *in vivo* paired recordings at thalamocortical synapses demonstrated that single axonal spiking can induce postsynaptic excitatory currents which can be detected extracellularly revealing a rise time constant around 0.3-1.5 ms and a decay time constant around 5-12 ms (Beierlein et al., 2003; Hull et al., 2009). In contrast, GABAergic evoked synaptic currents can generate extracellularly "unitary field potentials" that have the particular characteristic of being longer lasting (up to 5 ms duration, Glickfeld et al., 2009). Additionally, in sensory cortex *in vivo* paired patch clamp recording reported synaptically evoked GABAergic currents to be more prominent primarily during UP states, again with longer-lasting depolarizations (Jouhanneau et al., 2018, Sup. Fig. 6-7). These observations raise the possibility that longer-duration extracellular waveforms measured *in vivo* are originating from GABAergic inputs. (Telenczuk et al., 2017; Moore et al., 2017; Destexhe and Mehta, 2022). Notably, such waveforms can occur at rates largely exceeding neighboring somatic spiking (Destexhe and Mehta, 2022).

In summary, synaptically driven dendritic responses likely contribute to a spectrum of extracellular waveforms, both from AMPARs current which are short (1-3 ms long), and GABARs currents which are slightly longer (3-10 ms long) and of an opposite polarity.

## Dendritic spikes: calcium, NMDA and other longer-lasting plateaus

$Ca^{2+}$-dependent dendritic spikes are often initiated in distal dendritic regions, such as the apical tufts where $Ca^{2+}$ channel density is elevated. These events are broader in duration, involving significant $Ca^{2+}$ influx through various types of voltage gated $Ca^{2+}$ channels (e.g., L-type, T-type, R-type, N-type), which induce broader duration and lower amplitude depolarization relative to $Na^+$ spikes (rev. Stuart et al., 2015; Larkum et al., 2022). $Ca^{2+}$ influx through these channels also functions as an important second messenger, thereby linking electrical activity with intracellular signaling cascades implicated in synaptic plasticity, gene expression if not larger activation of dendrites. In addition, NMDA spikes (or NMDA dependent dendritic plateau potentials) are another type of nonlinear events (Larkum et al., 2001; 2007). Theycan last for tens to hundreds of milliseconds, providing sustained depolarization capable of modulating somatic output. Furthermore, the NMDARs function as coincidence detectors, requiring both glutamate binding and postsynaptic depolarization for significant current flow rendering them central to synaptic plasticity and dendritic computation (Hunt and Castillo, 2012; Lau and Zukin, 2007). Overall, it has been proposed that dendritic spikes are computationally efficient in strengthening fast event transmission, meanwhile opening longer windows to provide longer integrations (Burger et al., 2025).

Experimentally, a substantial literature has characterized dendritic spikes both *in vitro* and *in vivo* (rev. Larkum et al., 2022). One key property is their long duration (5 to 500 ms) and weak coupling to immediate somatic spiking (Larkum et al., 2007). These properties have implicated dendritic spikes in diverse brain processes such as anesthesia (Susuki et al., 2020), behavior (Takahashi et al., 2016; Takahashi et al., 2020), and sleep (Aime et al., 2022). Furthermore, these numerous, very active phenomena, occurring in dendrites, are known to consequently mediate drastic changes in the somatic firing patterns. For example, dendritic plateaus are known to be followed by somatic bursts of action potential in the cortex (Larkum et al., 1999) and the hippocampus (Bittner et al., 2017). Nevertheless, despite their importance and repeated experimental observations, there is to our knowledge no report of extracellular detections (or waveforms shapes) stemming from dendritic spikes, likely because their longer lasting duration >5 ms to 500 ms hinders direct detections by classical spikes sorting algorithms. So far, using paired recordings combining patch clamp of the soma and extracellular recording seems the best method of choice to measure these long-lasting events (Helmchen et al., 1999; Suzuki et al., 2017).

Here, we emphasize the possibility of systematically incorporating synaptically driven dendritic currents into extracellular signals analysis. In particular, AMPARs current kinetics have typical characteristics of being negative and lasting around 2 ms long. GABARs currents and kinetics are also very easily identifiable as being longer (3 - 10 ms) and of positive polarity. We encourage the field to further ask whether it would be possible to detect dendritic spikes extracellularly. This would be of prime importance knowing their critical influence on both neuronal encoding and behavior. Synaptic current kinetics and dendritic spikes events should be distinguished as longer lasting compared to bAPs and dendritic sodium spikes whose properties still remain within the scope of detectability and extractability given our current spike sorting methods. It is important to highlight that both ionotropic synaptic current, together with dendritic spikes all have an expected larger spatial footprint compared to somatic spikes (Fig. 4).

## Part V. Known combinations of compartment waveforms

Biophysical modelling has long predicted that the activities from multiple neuronal compartments summate in the extracellular space, generating waveforms with diverse and complex shapes (Gold et al., 2007; Zang et al., 2023). However, we advocate for a reduced outlook focusing on the known sequential activation of neuronal compartments rather than an exhaustive screening of all putative waveform combinations. In the following sections, we discuss the experimentally supported cases of waveforms combinations as in axonal volley and synaptic contacts.

### Synchronous axonal volley related activations

Axonal volleys have been most extensively characterized in the sensory systems (Wever and Bray 1930; Boudreau, 1965; Tasaki et al., 1983; rev. Curthoys et al., 2021). Crucially, they do not reflect the extracellular waveform from a single axon but emerge from synchronous activations across several parallel axons (Kuokkanen et al., 2010; Telenczuk et al., 2015). Rather than single-axon waveforms, population-level summation gives rise to compound action potentials (rev. Curthoys et al., 2021) and/or oscillatory frequency-following potentials in and around the auditory nerve (Worden and Marsh, 1968), in the auditory system (Tasaki, 1954; rev. Snyder and Schneider, 1984), and in the vestibular system (Cazals et al., 1979). Within the space in which neuronal activity is sufficiently synchronous, the extracellular field can summate constructively, producing waveforms across a large radius (Kuokkanen et al., 2010; Lindén et al., 2011). In addition, the number of parallel axonal projections within an axonal bundle can lead to larger extracellular contributions (Schmidt and Knöschke, 2019). However, for the LFP signature to faithfully resemble the underlying single-axon waveform, temporal synchrony must be considerably shorter than the duration of the single axonal waveform. Otherwise, the obtained LFP signature may become distorted. Indeed, the stochasticity of even the most synchronized responses, e.g., in peripheral auditory neurons with sub-millisecond response time dispersion, can lead to a "response volley". The volley's waveform shape differs from that of individual axonal LFPs (Boudreau, 1965) and during repetitive activation produces oscillatory patterns, e.g. frequency-following potentials (Kuokkanen et al., 2010).

## Waveforms measurements stemming from synaptic contacts

Double-peaked waveforms consist of an initial short, symmetric negative peak followed by a second longer-lasting negative peak. To explain preceding experimental findings (below), such waveforms were initially modelled as a possible summation of currents from different axonal branches of a single neuron (Gold et al., 2007). More recent biophysical modelling has demonstrated that these stereotypical waveforms may also correspond to excitatory synaptic contacts, where the first symmetric peak arises from axonal arbors and the second longer-lasting peak originates from synaptic currents (Hagen et al., 2017).

Experimentally, double-peaked waveforms, interpreted as an axonal arbor response followed by a postsynaptic response, were initially reported using paired recordings in the thalamocortical pathway (Swadlow and Gusev, 2000; Bereshpolova et al., 2019), the retinotectal pathway (Sauvé et al., 1995), and the corticotectal pathway (Su et al., 2024). Several recent studies have demonstrated that such double-peaked waveforms can be detected extracellularly using silicon probe *in vivo* in the thalamocortical pathway (Su et al., 2021; Sibille et al., 2024) and retinotectal pathway (Sibille et al., 2022; Gehr et al., 2023). The interpretation that the first negative peak accounts for the axonal arbor's waveform, and the second longer-lasting negative peak reflects the excitatory postsynaptic origin has been confirmed pharmacologically, both with silicon probe (Swadlow and Gusev, 2000) and high-density silicon probe (Sibille et al., 2022). Importantly, the kinetics of the second peak closely match the activation (0.2-0.8 ms) and deactivation (1.3-2.0 ms) time constants of AMPARs (Kleppe et al., 1999). Such an observation suggests that extracellular detections of double negative peaks may primarily reflect excitatory synaptic contacts, rather than downstream dendritic integration, which remains speculative.

Taken together, these converging lines of evidence highlight the existence of extracellular signals arising from tightly coupled presynaptic and postsynaptic processes, likely enabled by the dense clustering of close-by synapses activated nearly simultaneously by a single axon. Other studies have shown that these synaptically driven extracellular responses exhibit short-term plasticity, including facilitation (Teh et al., 2023) and depression (Swadlow et al., 2002), depending on the synapse type and pathway identity. Overall, given the ubiquity of excitatory synaptic contacts throughout neuronal tissues, these synaptically generated extracellular signals are likely common in *in vivo* recordings. By contrast, it remains an open question whether inhibitory synaptic contacts generate analogous extracellular signatures *in vivo*, with a negative short peak followed by a longer positive peak as reported in the turtle brain (Shein-Idelson et al., 2017), and observed in the hippocampus (Glickfeld et al., 2009).

## Amplitude, duration, and spatial footprint: a way to hierarchize extracellular signals from soma to dendrites?

In the previous paragraphs, we highlighted several waveform features that are rarely exploited systematically, most notably the spatial extent, or "spatial footprint" of extracellular waveforms. When combined with the duration of the different waveforms, spatial footprint can provide a clearer framework for organizing extracellular signals according to their compartment origin (Fig. 4). Waveform duration follows a clear gradient, ranging from short axonal waveforms to longer dendritic synaptic responses and plateau potentials (Fig. 4 D-E). Consequently, we propose that amplitude, duration, and spatial footprint be used more systematically in future extracellular detection and classification methods to enable more inclusive and compartment-aware interpretations.

Nevertheless, amplitude remains a dominant discriminative feature, as biophysical considerations link extracellular amplitude directly to the local density and synchrony of ionotropic channel activations. Within this framework, somatic spikes at the AIS produce the largest extracellular amplitudes of all waveforms (Bakkum et al., 2019). The next largest contribution likely arises from axonal arbors and nodes of Ranvier, although this effectively constrains axonal detection to long-range spatially organized projections. Dendritic synaptically evoked currents are expected to produce intermediate amplitudes, while single axonal projections and passive or active dendritic responses generate the smallest extracellular signals. We summarize these relationships, together with spatial extent (Fig. 4F) and duration, in a simplified hierarchical framework (Fig. 4), which we propose as a practical guide for interpreting complex extracellular waveforms. For simplicity, we have here neglected both the state-dependent variability in waveforms (with brain state, anesthesia, behavior, firing frequency, etc.) and how tissue inhomogeneities, anisotropy, and frequency-dependent filtering affect recordings.

## Conclusion

In conclusion, waveform duration and spatial footprint emerge as particularly powerful and underexploited features for advancing three-dimensional template matching approaches. We highlight these features as possibly determinant in distinguishing extracellular waveforms originating from the three principal neural compartment waveforms. Somatic spike waveforms have medium to large amplitudes (>50 µV), intermediate spatial footprints (∼150 µm), and durations of 0.3-0.8 ms. Axonal waveforms, by contrast, are shorter in duration (~0.1-0.3 ms) and exhibit a wide range of spatial footprints, ranging from very small footprint for nodes of Ranvier (<20 µm) to very large for certain axonal arbors (>400 µm). Dendritic waveforms are typically longer in duration (5-500 ms), and therefore largely overlooked by current spike-sorting methods, with a few exceptions. In particular, bAPs have the same shape as the somatic spikes but exhibit a progressive fading amplitude with increasing distance from the soma. AMPA-mediated synaptic currents are longer (~1-3 ms long) and display relatively large spatial footprints (>250 µm). GABA-mediated synaptic currents are even longer (~3-10 ms long) and are also expected to exhibit large spatial footprints (>250 µm), although this has yet to be shown in *in vivo* recordings. Altogether, these biophysical predictions, supported by converging experimental observations, pave the way for a more inclusive mechanistic understanding of the electrophysiological extracellular signals generated in neural tissues.

Two additional layers of complexity must be considered when advancing any claim toward an inclusive three-neuronal-compartments classification framework. First, close morphological proximity of the electrically isolated neuronal compartments inevitably leads to overlap and summation of extracellular waveforms, as elegantly exemplified in the double-peaked synaptic waveforms and extensively quantified in the axonal volley. Second, temporal synchrony across neurons represents a fundamental challenge, as neuronal computation inherently involves the concomitant activation of neighboring neurons. This makes the extracellular detection and identification of individual neuronal compartments challenging.

To conclude, on a forward-looking note, we identify three avenues for future methodological and conceptual progress. First, current spike-sorting frameworks should explicitly integrate axonal action potentials and synaptic responses, along with classical extracellular somatic action potentials, possibly by expanding and refining more extensive template libraries of time-dependent waveforms to include the three principal neuronal compartments. Second, template-matching algorithms should integrate spatial considerations more systematically, including spatial footprint and the propagation directionality of compartment-specific signals such as axonal projections, dendritic downstream signals, and bAPs. The widespread use of high-density silicon probes provides an unprecedented opportunity to incorporate such spatial considerations, which should be leveraged to our advantage. Third, template-matching approaches, including deep-learning-based methods, may be revisited to better handle temporal synchrony and waveform overlap. This will likely require an enlargement of the analysis window duration used in spike sorting, while carefully screening all possible combinations from the three principal compartments in a single pass.

Box 1: Feasibility of detecting signals with low amplitude and large spatial reach

In today's electrophysiological equipment the impedance -as of the input resistance- is usually low (<1 MΩ) compared to earlier technologies such as extracellular pipette or tungsten electrodes (~10 MΩ). Furthermore, impedance adaptation is now routinely done in high density electrodes (Jun et al., 2016; Steinmetz et al., 2021; Ye et al., 2025) bringing the virtual input resistance down to ~150 kΩ on almost all channels (Lopez et al., 2017). Consequently, one can approximate the spatial extent of normal somatic waveforms in the mouse cortex to be around 200 µm (rev. Buzsaki, 2004), confirmed by looking into spike triggered LFPs (Sibille et al., 2022; Sibille et al., 2024). Quite interestingly, lead by the Kempter lab for a few years, single spikes have been revealed to be measurable in the EEG signal using similar spike triggered LFP (Telenczuk et al., 2015; Kuokkanen et al., 2018, 2025) literally reconstructing single cell's spikes measured by EEG almost 10 mm away for its source (Kuokkanen et al., 2025). For such measure, the obtained spike-triggered LFP range of nano-volts, which is a signal way too small to be detected from EEG without extensive averaging over thousands of spikes to eliminate the unrelated signals. Nevertheless, the changes of detectability of single events from tetrodes to high-density silicon probe in benefiting from the high spatial density which can more properly unravel neuronal compartments beyond somatic spikes, as already observed using more classical silicone probes or tetrodes (Sun et al., 2023; Someck et al., 2023; Senzai et al., 2019).

Box 2: Grounding is a must to record smaller signals

Grounding and referencing are two separate concepts that require dedicated attention; especially when aiming for low-noise level recordings tuned for neural compartment waveforms measured beyond somatic waveforms.

The reference is the signal every other signal(s) is(are) subtracted to, usually the screw in the cerebellum for freely moving recording conditions or the bath in the crown for head-fixed acute preparation. Somehow this subtraction is made to remove any signal being similar in both the reference channels and your recordings channel(s) as obviously the noise produced by surrounding elements, such as body movements, licking apparatus, magnetic field from earth, changes of electric power in the wall etc. The reference consequently removes spatially large signals being similar to both recording and reference channels. Therefore, any other more localized signal, the power supply of a screen, the light bulb in your room etc. are all sources of noise which are more localized and therefore different in both the reference and recording channel(s).

Practically, in physics, one can use the analogy of comparing electricity as the oscillation that spreads through the surface of water, to make problems easier to be visualized. Using this view, any electrical equipment anywhere around a set-up (including a big fridge on the other side of the wall -even turned off but plugged to its power socket-) should be viewed as an arm tapping onto the water surface. Then the solution to "break the tapping arm" is to ground a point between the arm and your recording. A faraday cage, an aluminum foil, a metal mesh, even your colleague will do this work just fine when properly grounded. Consequently, one needs to answer three points to do proper grounding: 1- Where to find a proper ground contact? 2- Where/what is the source of noise? 3- How to individually ground it?

1- Grounding points are available in every power socket as the outer pin of power supplies (or the side metal hook). In certain buildings, an extra thick copper cable is sometimes available in the experimental room (often covered in a typical yellow with green strip colored plastic), which you supposedly should also have in the closest electrical panel around you.

2- You probably have heard it, as most electrophysiologists know: touching some part of your set-up (or head stage) with your naked finger can temporarily reduce the noise. If this magical trick works, it is because you are acting as a local grounding point (as we are mostly composed of water), exactly as when putting your finger on the water surface in front of the tapping arm, by reducing locally and temporarily the undergoing oscillation. So indeed, it is a way to look at the electrophysiological signal you aim to record from meanwhile touching different parts of the surrounding set-up/recording system. This can help in identifying the source(s) of noise. Alternatively, you can use the naked probe and move it around to evaluate whether the left or the right side of the room gets more noise. To be very thorough, don't hesitate to fix your recording elements in a little PBS bath (which should have the reference channel in) and place it in the different corners of your experimental room. Doing so you should be able to empirically see where what for noise is. If these tricks do not work, then the last resort will be to process by elimination in removing all equipment from the recording environment one piece at a time, until you identify this noise source bugging your recording.

3- Once noisy sources are identified, covering the element such with a foil, a copper mesh, a metal plate, which is directly connected toward your closest grounding point should isolate your recording from this noise source. It is important to avoid multiple parallel connections which can produce loops. If multiple connections are required using a spider web where its central point is the grounding point on the wall will avoid any extra noise. In the case of a faraday cage, it is wise to verify whether the different sides are properly connected to each other. Do not be afraid to have 3 to 5 different separate points of your faraday cage being connected to your grounding points in the wall. For recordings in freely moving animals, the experimental room can be seen as a faraday cage, where having metal plates on the sides, or on the room floor, should equally act as a classical faraday cage. Ideally, as done in certain universities, the walls of the room have been pre-covered with copper mesh behind the plaster, all connected to an available grounding point.

Importantly, noise can be taken care of "once for all", if all noise sources are properly identified. Some difficult cases are often due to power supply cables or requiring a change of a piece of equipment.

# Acknowledgments

We would like to acknowledge the entire Schmitz lab, the Kremkow lab, the Kempter lab for providing administrative and license support. We would like to acknowledge the great work supported by the Neuropixels community, and the underlying international collaboration sustaining such high-tech as production costs.

# Author contributions:

All authors listed have made direct, substantial and intellectual contributions to the work, and approved it for publication.

Conception and design: J. S.
Acquisition of data: not relevant
Analysis and interpretation of data: K. T., P. K., A. T., J. S.
Drafting or revising the article: J. S., K. T., A. T., D. S., P. K.

# Fundings

This study was supported by the German Research Foundation (Deutsche Forschungsgemeinschaft (DFG), grant nr. 502188599, project 184695641 – SFB 958, project 327654276 – SFB 1315, project 415914819 – FOR 3004, project 431572356, and under Germany's Excellence Strategy – Exc-2049-390688087), by the Federal Ministry of Education and Research (BMBF, SmartAge-project 01GQ1420B) and by the European Research Council (ERC) under the Europeans Union's Horizon 2020 research and innovation program (grant agreement No. 810580).